\documentclass[
]{ceurart}

\usepackage{listings}
\usepackage{graphicx}
\usepackage{tikz}
\usetikzlibrary{bpmn, positioning, fit, arrows.meta, backgrounds}
\usepackage{float}

\definecolor{dutyblue}{HTML}{BBDEFB}
\definecolor{cdblue}{HTML}{90CAF9}
\definecolor{cpgreen}{HTML}{C8E6C9}
\definecolor{probred}{HTML}{FFCDD2}

\begin{document}

\copyrightyear{2026}
\copyrightclause{Copyright for this paper by its authors.
  Use permitted under Creative Commons License Attribution 4.0
  International (CC BY 4.0).}

\conference{ODRL and beyond: practical applications and challenges for policy-base access and usage control.
OPAL 2026, 2nd edition
Co-located with the Extended Semantic Web Conference
Dubrovnik , Croatia · 10th-11th MAY (half-day), 2026}

\title{Extracting ODRL Policies from Business Process Models: A Graph Traversal Approach to Compliance-by-Extraction}

\author[1]{Meem Arafat Manab}[%
orcid=0000-0002-2336-4160,
email=meem.manab@upm.es,
]
\cormark[1]
\address[1]{Ontology Engineering Group, Departamento de Inteligencia Artificial,
  Universidad Polit\'{e}cnica de Madrid, Campus de Montegancedo,
  28223 Boadilla del Monte, Madrid, Spain}

\author[2,3]{Marinella Quaranta}[%
orcid=0000-0003-2691-0611,
]
\address[2]{Dipartimento di Informatica -- Scienza e Ingegneria (DISI),
  Alma Mater Studiorum -- Universit\`{a} di Bologna,
  Viale del Risorgimento 2, 40136 Bologna, Italy}
\address[3]{Dipartimento di Informatica, Universit\`{a} degli Studi di Torino,
  Corso Svizzera 185, 10149 Turin, Italy}

\author[3]{Ilaria Angela Amantea}[%
orcid=0000-0003-1329-1858,
]

\author[1]{Sheyla Leyva-S\'{a}nchez}[%
orcid=0009-0007-9762-4045,
]

\author[1]{V\'{i}ctor Rodr\'{i}guez-Doncel}[%
orcid=0000-0003-1076-2511,
]

\cortext[1]{Corresponding author.}
\begin{abstract}
Organisations maintain large corpora of process models expressed in the Business Process Model and Notation (BPMN), yet the normative content encoded in those models, the obligations, permissions, and prohibitions that govern participant behaviour, remains inaccessible to policy infrastructure. The Open Digital Rights Language (ODRL) is the emerging lingua franca of machine-readable policy, but authoring ODRL at scale is slow, expert-intensive work, and generation by large language models introduces well-documented risks of structural invalidity. We present a pipeline that resolves this gap by extracting ODRL policies automatically from BPMN XML, grounded in the observation that BPMN control flow encodes deontic modalities by construction. The pipeline traverses the BPMN process graph, classifies each task as an \texttt{odrl:Duty} or \texttt{odrl:Permission} via a reachability check, and introduces a novel treatment of intermediate catch events as \texttt{odrl:Prohibition} rules with lifting constraints, capturing waiting semantics that prior approaches have dropped. The result is a compliance-by-extraction approach in which auditable, interoperable ODRL policies are derived directly from the process models organisations already maintain. The implementation and a live demonstration are available at \url{https://github.com/manabcodes/bpmn2odrl} and \url{https://bpmn.linkeddata.es/}.
\end{abstract}


\maketitle

\section{Introduction}

Organisations have spent decades building and maintaining process models. These models, expressed in the Business Process Model and Notation (BPMN)~\cite{omg2013bpmn}, are operational documentation: they describe how goods get dispatched, how loans get approved, how medical referrals get processed. They are not, in intent, compliance artefacts. But they are, in structure, normative ones. Every well-formed BPMN process encodes, as a direct consequence of its control flow, a set of obligations, permissions, and prohibitions: which participant must perform which task, which tasks may be skipped, and which roles are excluded from which actions entirely. This normative content is not an annotation layer added on top of the process; it is constitutive of the process structure itself, a point established formally by Natschl\"{a}ger~\cite{natschlager2011deontic}.

The Open Digital Rights Language (ODRL)~\cite{odrl2018}, a W3C Recommendation with a growing ecosystem of compliance profiles targeting the General Data Protection Regulation (GDPR), the EU AI Act, and data space governance frameworks such as Gaia-X and the International Data Spaces (IDS), is the emerging lingua franca of machine-readable policy. Its primitive vocabulary, \texttt{odrl:Duty}, \texttt{odrl:Permission}, and \texttt{odrl:Prohibition}, maps directly onto the deontic modalities that Natschl\"{a}ger proves are structurally present in BPMN, making it the natural and principled extraction target. ODRL is increasingly referenced and adopted across international initiatives and governance frameworks. It serves as the default policy language in the JPEG Trust standard, is incorporated into the International Data Spaces Association Reference Architecture Model and its Usage Control Language, and has been examined by the Data Spaces Support Centre and Gaia-X for expressing access and usage control policies. Additionally, the European Union Intellectual Property Office recommends ODRL for representing machine-readable rights and obligations in open rights data exchanges. Policies generated by large language models are an increasingly common response to this shortage. However, Mustafa et al.~\cite{mustafa2024odrl} show that LLMs generating ODRL policies substitute undefined vocabulary terms for standard ones, using, for instance, \texttt{odrl:location} in place of the correct \texttt{odrl:spatial}, a deviation that renders the policy uninterpretable by any conformant ODRL processor despite appearing semantically reasonable. Policies extracted algorithmically from verified process models do not carry this risk.

This is precisely the gap that the present paper addresses. BPMN is operationally useful but semantically opaque to policy infrastructure: a compliance checker, a data space broker, or a regulatory audit tool cannot reason over a BPMN XML file in terms of obligations, permissions, and prohibitions
. The normative content those tools require is already present in the BPMN structurally, by construction, not by annotation, but it is locked in a form that policy infrastructure cannot consume. The warehouse BPMN already encodes that the warehouse worker must verify the order before shipping; that obligation does not need to be written, it needs to be extracted. We present a pipeline that performs this extraction automatically, traversing the BPMN process graph and producing valid ODRL policies in JSON-LD, turning an existing corpus of billions of process models into a source of machine-readable compliance artefacts without any additional authoring effort.

The practical significance of this approach extends beyond the supply problem. The conventional route to compliance is assertion: an organisation authors a policy document, or an ODRL file, claiming conformance with some requirement. The pipeline we describe offers something stronger, which we term compliance-by-extraction: the policy is derived algorithmically from the process that is actually being executed. If the process changes, the extracted policy changes with it. The resulting artefact is traceable back to its source, producing an auditable chain from ODRL rule to BPMN task to production process, of a kind that authored or generated policies cannot provide.

The remainder of this paper is structured as follows. Section~\ref{sec:background} situates the work with respect to ODRL, deontic BPMN, and compliance-by-design. Section~\ref{sec:pipeline} presents the mapping from BPMN constructs to ODRL primitives and the extraction algorithm. Section~\ref{sec:evaluation} demonstrates the pipeline on the credit scoring benchmark process. Section~\ref{sec:limitations} discusses open problems, and Section~\ref{sec:conclusion} concludes our discussion.

\noindent The contributions of this paper are:
\begin{itemize}
\item A graph traversal pipeline that extracts valid ODRL policies in JSON-LD directly from BPMN XML, requiring no manual policy authoring.
\item A reachability-based classification of BPMN tasks as \texttt{odrl:Duty} or \texttt{odrl:Permission}, rooted in Natschl\"{a}ger's formal deontic BPMN result, with a corrected treatment of parallel gateways.
\item A novel mapping of intermediate catch events to \texttt{odrl:Prohibition} rules with lifting constraints, capturing waiting semantics that prior approaches have discarded.
\item A compliance-by-extraction framing in which policies are derived from the process actually being executed, producing an auditable chain from ODRL rule to BPMN task.
\end{itemize}

\section{Background}
\label{sec:background}

Three bodies of work converge to motivate the pipeline presented in this paper: the formal study of deontic content in BPMN, the compliance-checking research programme that has treated BPMN and norms as separate artefacts to be reconciled, and the growing ecosystem of ODRL profiles that has established the language as the standard vehicle for machine-readable regulatory policy.

\subsection*{Deontic content in BPMN}

The observation that business process models carry normative content is not new. Governatori and Sadiq~\cite{governatori2009journey} established the compliance problem as a relationship between two distinct specification sets: the rules governing how a
process executes and the norms that govern how it must behave. The dominant research response has been to treat these sets as given separately and to check one against the other at design time or at runtime~\cite{hashmi2018we}. A substantial body of work by Governatori and colleagues formalises this checking problem using defeasible deontic logic, providing algorithms that traverse a process graph, accumulate task effects, and determine whether obligations derived from regulations are fulfilled, violated, or compensated~\cite{governatori2008detecting}. That programme assumes, however, that the normative content to be checked has already been authored as an explicit rule set. The question of whether normative content is already structurally present inside the process model itself is a different question, and it was answered formally by Natschl\"{a}ger~\cite{natschlager2011deontic,natschlager2015deontic}. Working from the observation that BPMN control flow constrains which tasks can be skipped and which cannot, Natschl\"{a}ger demonstrates that tasks on every execution path imply obligations, tasks reachable only through an exclusive gateway branch imply permissions, and the absence of reachability from a given role implies a prohibition. The result is proven by a graph transformation system shown to be terminating, confluent, and behaviour-preserving. This is the formal foundation on which the present paper rests: the deontic content is not added to a BPMN model, it is extracted from one.

\subsection*{BPMN formalisation and process semantics}

Algorithmic extraction of any kind over BPMN requires that the notation carry a well-defined formal semantics. Dijkman, Dumas and Ouyang~\cite{dijkman2008semantics} provide this through a modular mapping of BPMN constructs to Petri nets, covering tasks, all gateway types, events, and sub-processes, and identifying deficiencies in the standard's own informal characterisation of its constructs. This Petri-net semantics establishes that BPMN process graphs are amenable to static analysis, including the reachability checks on which the present pipeline relies for duty-versus-permission classification. Subsequent work has explored richer formalisation through ontologisation: Annane, Aussenac-Gilles and Kamel~\cite{annane2019bbo} present BBO, an OWL~2 DL ontology covering BPMN~2.0 elements with reasoning support and SPARQL queryability over instantiated process models. Di Martino et al.~\cite{di2023tool} demonstrate semantic annotation and Prolog-based validation of BPMN models for regulatory purposes. Neither approach produces output in a standardised policy language, and neither operates without manual annotation effort.

\subsection*{ODRL as a compliance policy language}

The Open Digital Rights Language~\cite{odrl2018} was designed for digital rights management but its primitive vocabulary, \texttt{odrl:Permission}, \texttt{odrl:Prohibition}, and \texttt{odrl:Duty}, maps cleanly onto the deontic modalities that Natschl\"{a}ger demonstrates are present in BPMN by construction. De Vos, Kirrane, Padget and Satoh~\cite{de2019odrl} first demonstrated that ODRL can represent regulatory obligations and permissions beyond the DRM domain, introducing the ODRL Regulatory Compliance Profile (ORCP) and a compliance-checking pipeline that translates ODRL policies into Answer Set Programming for automated verification against GDPR fragments. That paper establishes ODRL as a viable compliance language but requires all policies to be authored manually. The supply problem this creates has grown more acute as ODRL has been adopted as a normative component of major European data infrastructure. The International Data Spaces Association mandates ODRL for all usage-control policies in the Dataspace Protocol~\cite{idsa2021}, and Gaia-X adopts the ODRL information model as its generic policy representation~\cite{gaiax2024}. Golpayegani et al.~\cite{golpayegani2024aiup} introduce AIUP, an ODRL profile for expressing intended and precluded uses of AI systems under the EU AI Act. ODRL is advancing toward ISO standardisation, and the breadth of profiles now targeting GDPR, the EU AI Act, and data space governance collectively confirm it as the reference language for machine-readable regulatory policy in the European data ecosystem.

\subsection*{The gap}

The compliance-checking literature takes norms and process models as two independently authored artefacts and asks whether the second satisfies the first. Natschl\"{a}ger shows that deontic modalities are structurally present in the second artefact already, but maps them to a custom notation rather than to a standardised policy language. ODRL is now that standardised language, adopted normatively by multiple regulatory and governance frameworks, but its policies must still be written by hand, a bottleneck that Hashmi et al.~\cite{hashmi2018we} had identified as a persistent open problem in the compliance literature. The present paper closes this triangle: it extracts ODRL policies automatically from BPMN models, grounding the extraction in Natschl\"{a}ger's formal result and producing artefacts that are immediately consumable by the ODRL ecosystem without any manual policy-authoring step.

\section{Mapping and Pipeline}
\label{sec:pipeline}

The extraction rests on a structural correspondence between BPMN constructs and ODRL primitives. Table~\ref{tab:mapping} states this correspondence formally. The correspondence is not an encoding choice; it follows from the deontic semantics of BPMN control flow as established by Natschl\"{a}ger~\cite{natschlager2011deontic}, with two departures described below.

\begin{table}[h]
\centering
\caption{BPMN to ODRL mapping}
\label{tab:mapping}
\begin{tabular}{lll}
\toprule
\textbf{BPMN construct} & \textbf{ODRL primitive} & \textbf{Condition} \\
\midrule
Participant / pool        & \texttt{odrl:Party}         & Assignee of all rules in that pool \\
Task on every path        & \texttt{odrl:Duty}          & Removal disconnects all end states \\
Task on every path within a branch & \texttt{odrl:Duty} + constraint & Branch itself is optional \\
Task on at least one skippable path & \texttt{odrl:Permission}   & At least one end state survives removal \\
XOR gateway branch        & \texttt{odrl:Permission}    & Exactly one branch taken \\
AND gateway branch        & \texttt{odrl:Duty}          & All branches mandatory; no choice exists \\
EVENT gateway branch      & \texttt{odrl:Permission}    & Conditioned on which event fires \\
Intermediate catch event  & \texttt{odrl:Prohibition}   & Lifted when triggering event fires \\
Cross-role task           & \texttt{odrl:Prohibition}   & Closed-world: task belongs to another pool \\
Outgoing message flow     & \texttt{odrl:Duty} (inform) & Notify obligation on sending task \\
Swim lane assignment      & \texttt{odrl:assignee}      & Party constraint on the rule \\
\bottomrule
\end{tabular}
\end{table}

We depart from Natschl\"{a}ger on two points. First, she reads the parallel gateway split as permission, on the grounds that it introduces branching. We do not. A
parallel gateway forces all branches to execute; the participant has no choice about which branches to take or whether to take them, and where there is no choice, there is no permission. Each parallel branch is classified as an \texttt{odrl:Duty}. Second, and more substantially, Natschl\"{a}ger does not address intermediate catch events. Prior work has treated them as transparent pass-through nodes, dropping their normative content or attaching it as an ad hoc constraint on downstream tasks. We argue that neither treatment is correct. An intermediate catch event encodes a conditional prohibition on proceeding: the participant is forbidden from executing the downstream task until the triggering condition is satisfied. This is not a duty, because waiting is not an action the participant performs, and it is not a permission, because the participant has no choice about waiting. It is a prohibition on continuation, lifted when the event fires, and it maps onto \texttt{odrl:Prohibition} with a constraint encoding the lifting condition.


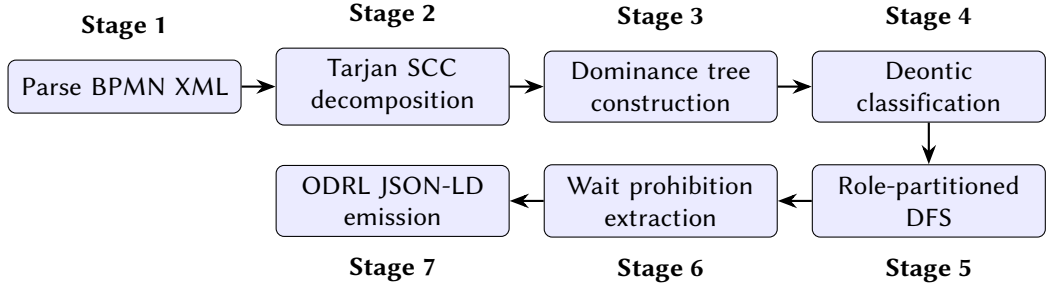
\begin{figure}[h!]
\centering
\begin{tikzpicture}[
  font=\small,
  box/.style={rectangle, draw, rounded corners=3pt,
              text width=2.8cm, minimum height=0.7cm,
              inner sep=4pt, align=center,
              fill=blue!8},
  arr/.style={-{Stealth[scale=1.1]}, thick},
  lbl/.style={font=\small\bfseries, text=black}
]

\node [box] (s1) {Parse BPMN XML};
\node [box, right=0.45cm of s1] (s2) {Tarjan SCC \\ decomposition};
\node [box, right=0.45cm of s2] (s3) {Dominance tree \\ construction};
\node [box, right=0.45cm of s3] (s4) {Deontic \\ classification};
\node [box, below=0.55cm of s4] (s5) {Role-partitioned \\ DFS};
\node [box, left=0.45cm of s5] (s6) {Wait prohibition \\ extraction};
\node [box, left=0.45cm of s6] (s7) {ODRL JSON-LD \\ emission};

\foreach \s/\t in {s1/s2, s2/s3, s3/s4, s4/s5, s5/s6, s6/s7}
  \draw [arr] (\s) -- (\t);

\node [lbl, above=0.15cm of s1] {Stage 1};
\node [lbl, above=0.15cm of s2] {Stage 2};
\node [lbl, above=0.15cm of s3] {Stage 3};
\node [lbl, above=0.15cm of s4] {Stage 4};
\node [lbl, below=0.15cm of s5] {Stage 5};
\node [lbl, below=0.15cm of s6] {Stage 6};
\node [lbl, below=0.15cm of s7] {Stage 7};

\end{tikzpicture}
\caption{The seven stages of the BPMN-to-ODRL extraction pipeline.}
\label{fig:pipeline}
\end{figure}

The pipeline implements this mapping as a seven-stage graph traversal over the BPMN process graph,  illustrated in Figure~\ref{fig:pipeline}. Stage~1 parses the BPMN XML into a directed graph of tasks, gateways, and events, extracting swim-lane assignments, sequence flow conditions, and message flow directions. Well-formedness is enforced by Python's XML parser, and then the input is validated against the BPMN~2.0 XML Schema~\cite{omgbpmn2010xsd} before graph construction. Malformed or schema-invalid documents are rejected with a parse error, and nodes lacking a required \texttt{id} attribute are silently skipped.  Party identity is derived from the \texttt{processRef} attribute of each \texttt{<participant>} element in the \texttt{<collaboration>} block, making the pool the unit of party assignment, and swimlane structures within a pool are parsed but not used for further party refinement.

 Stage~2 applies Tarjan's Strongly Connected Components algorithm~\cite{tarjan1972depth} to detect and collapse cycles into meta-nodes, producing a directed acyclic graph; ODRL has no loop construct, and cycles must be resolved before policy extraction can proceed. Stage~3 constructs the dominance tree using a fixed-point iteration over the acyclic graph~\cite{lengauer1979fast}, establishing precondition chains between tasks. Stage~4 classifies each task as duty or permission using a per-node reachability check: the task is removed from the graph, and a breadth-first search determines whether any end state remains reachable; if none does, the task is mandatory. A second pass within gateway branches refines classifications for tasks that are skippable globally but constrained within a branch: the first task encountered (by breadth-first order) on each exclusive-gateway branch is emitted as \texttt{odrl:Permission} with the branch condition as constraint, since it represents the entry point of an optional execution path; subsequent tasks on the same branch, which are mandatory given that the branch was entered, are emitted as \texttt{odrl:Duty} with the same constraint. A branch whose sole task is also its first task therefore always yields a \texttt{odrl:Permission}, as illustrated by \texttt{report delay} in Figure~\ref{fig:extraction}, and by \texttt{compute score L2}, whose branch also contains \texttt{send credit score}, classified as \texttt{odrl:Duty}, since it follows the branch-entry permission. Stage~5 walks all intermediate catch events and emits one \texttt{odrl:Prohibition} per downstream task, with a constraint encoding the lifting condition and, where a message flow identifies the sender, an additional party constraint. Stage~6 performs a path-aware depth-first traversal~\cite{tarjan1972depth} per pool, accumulating gateway conditions per task across all paths and unioning them; a task reachable unconditionally on any path is emitted unconditionally overall. Outgoing message flows are emitted as notify duties at this stage. Stage~7 assembles the final \texttt{odrl:Set} policy in JSON-LD, with one rule per classified task, gateway conditions as \texttt{odrl:constraint} instances, and one compact cross-role \texttt{odrl:Prohibition} per participant listing all tasks that belong exclusively to other pools. The pipeline is implemented in Python and released as open-source software; the repository, along with a live demonstration interface, is available at \url{https://anonymous.4open.science/r/bpmn2odrl-4F4B/} and \url{https://sage-lollipop-c59903.netlify.app} respectively.

\section{Evaluation}
\label{sec:evaluation}

\begin{figure}[p]
\centering
\includegraphics[width=1.0\linewidth]{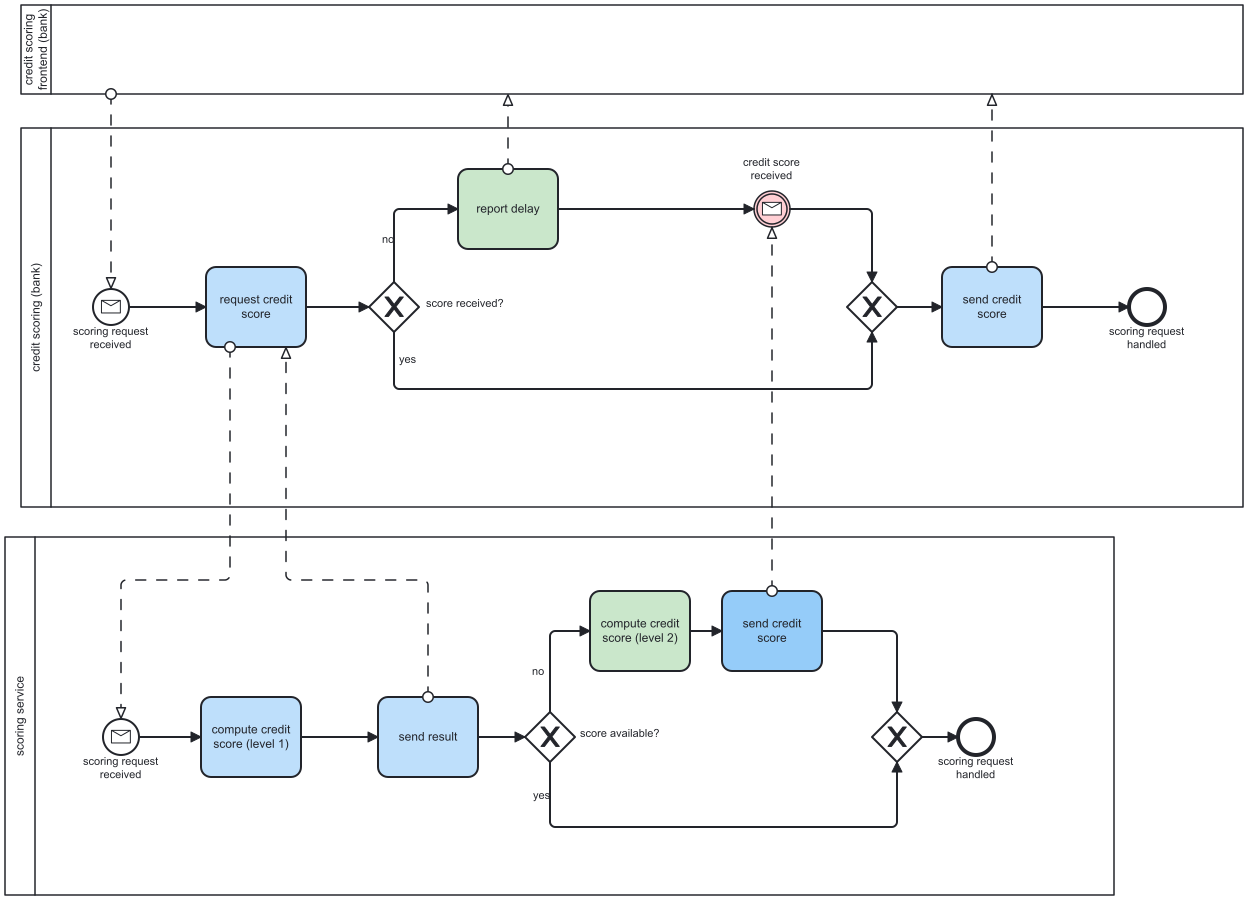}
\vspace{4pt}
\begin{tikzpicture}
  \draw[very thick, double distance=2pt]
    (0, 0.8) -- (0, 0.2);
  \node at (0, 0) {\textbf{Extraction using graph traversal algorithms}};
  \draw[-{Stealth[scale=1.4]}, very thick, double distance=2pt]
    (0, -0.3) -- (0, -1.2);
\end{tikzpicture}
\vspace{4pt}
\begin{tikzpicture}[
  font=\scriptsize,
  box/.style={rectangle, draw, rounded corners=2pt,
              text width=3.6cm, minimum height=0.75cm,
              inner sep=3pt, align=left},
]
\node [box, fill=dutyblue] (d1)
  {\textbf{odrl:Duty} \\ \textit{request credit score} \\ assignee: bank};
\node [box, fill=dutyblue, right=0.25cm of d1] (d2)
  {\textbf{odrl:Duty} \\ \textit{send credit score} \\ assignee: bank};
\node [box, fill=dutyblue, right=0.25cm of d2] (d3)
  {\textbf{odrl:Duty} \\ \textit{compute score L1} \\ assignee: scoring svc};
\node [box, fill=dutyblue, right=0.25cm of d3] (d4)
  {\textbf{odrl:Duty} \\ \textit{send result} \\ assignee: scoring svc};
\node [box, fill=cdblue, below=0.2cm of d1] (cd1)
  {\textbf{odrl:Duty} \\ \textit{send credit score} \\ \texttt{odrl:constraint}: \\\quad score avail.\ = no};
\node [box, fill=cpgreen, below=0.2cm of cd1] (cp1)
  {\textbf{odrl:Permission} \\ \textit{compute score L2} \\ \texttt{odrl:constraint}: \\\quad score avail.\ = no};
\node [box, fill=cpgreen, right=0.25cm of cp1] (cp2)
  {\textbf{odrl:Permission} \\ \textit{report delay} \\ \texttt{odrl:constraint}: \\\quad score received? = no};
\node [box, fill=probred, below=0.2cm of cp1] (wp1)
  {\textbf{odrl:Prohibition} \\ bank $\neg$ \textit{send credit score} \\ until: credit score received};
\node [box, fill=probred, below=0.2cm of wp1] (rp1)
  {\textbf{odrl:Prohibition} \\ bank $\neg$ compute L1/L2};
\node [box, fill=probred, right=0.25cm of rp1] (rp2)
  {\textbf{odrl:Prohibition} \\ scoring svc $\neg$ request \\ credit score};
\begin{scope}[on background layer]
  \node [draw=gray, rounded corners=3pt,
        fit=(d1)(d2)(d3)(d4)(cd1)(cp1)(cp2)(wp1)(rp1)(rp2),
         inner sep=6pt,
         label={[font=\scriptsize\bfseries]above:ODRL Policy
                (\texttt{odrl:Set})}] (frame) {};
\end{scope}
\matrix [below=0.3cm of frame.south west, anchor=north west,
         column sep=5pt, row sep=1pt, font=\tiny] {
  \node [fill=dutyblue,  draw, minimum size=7pt] {}; &
  \node {\texttt{odrl:Duty} (unconditional)}; &
  \node [fill=cdblue,    draw, minimum size=7pt] {}; &
  \node {\texttt{odrl:Duty} (constrained)}; \\
  \node [fill=cpgreen,   draw, minimum size=7pt] {}; &
  \node {\texttt{odrl:Permission} (constrained)}; &
  \node [fill=probred,   draw, minimum size=7pt] {}; &
  \node {\texttt{odrl:Prohibition}}; \\
};
\end{tikzpicture}

\caption{The synchronous credit scoring process (top) and the ODRL policy extracted from it (bottom). Colour coding links each rule to its type. Dashed arrows in the BPMN diagram are message flows between pools.}
\label{fig:extraction}
\end{figure}

We evaluate the pipeline on process models drawn from the Camunda \textit{bpmn-for-research} benchmark~\cite{camunda2015bpmnresearch}, covering four canonical scenarios: Dispatch of Goods, Recourse, Credit Scoring, and Self-Service Restaurant. Credit Scoring contributes two distinct types of models, a synchronous and an asynchronous variant; Figure~\ref{fig:extraction} illustrates the extracted policy for the synchronous case. Together the five models exercise the full range of constructs the pipeline is designed to handle: mandatory sequence paths, XOR and AND gateways, event-based gateways, intermediate catch events, multi-party message flows, and one cycle. Table~\ref{tab:results} summarises the output for each model.

\begin{table}[h]
\centering
\caption{Pipeline output across the five benchmark models}
\label{tab:results}
\footnotesize
\setlength{\tabcolsep}{4pt}
\begin{tabular}{lrrrrrrrrr}
\toprule
\textbf{Model} & \textbf{Nodes} & \textbf{Parties} & \textbf{D} &
\textbf{CD} & \textbf{P} & \textbf{CP} & \textbf{WP} & \textbf{RP} \\
\midrule
Dispatch of Goods       & 15 & 1 &  1 & 2 & 2 & 2 & 0 & 0 \\
Recourse                & 20 & 1 &  1 & 5 & 0 & 3 & 3 & 0 \\
Credit Scoring (sync.)  & 7  & 2 &  4 & 1 & 0 & 2 & 1 & 2 \\
Credit Scoring (async.) & 19 & 3 &  3 & 2 & 0 & 3 & 3 & 2 \\
Self-Service Restaurant & 30 & 3 & 18 & 0 & 0 & 0 & 5 & 3 \\
\bottomrule
\end{tabular}
\par\smallskip
\raggedright\footnotesize
D = Duty, CD = ConstrainedDuty, P = Permission,
CP = ConstrainedPermission, WP = WaitProhibition, RP = RoleProhibition
\end{table}
The pipeline produces valid \texttt{odrl:Set} policies in JSON-LD for all five models without error. We discuss each model in turn.

\textit{Dispatch of Goods} is a single-participant process with an inclusive gateway, making it the most structurally constrained of the five models. The pipeline correctly identifies one unconditional duty (preparing goods for pickup), two constrained duties within the special handling branch, and four permissions, two of which carry gateway conditions. No intermediate catch events are present, and no cross-role prohibitions arise from a single-pool process.

\textit{Recourse} is also a single-participant process but contains three intermediate catch events, making it the primary test case for the wait-as-prohibition treatment. The pipeline emits three \texttt{odrl:Prohibition} rules, one for each catch event, correctly identifying that the clerk is prohibited from proceeding to \textit{check reasoning} until a disagreement letter is received, from proceeding to \textit{hand over to collection agency} until a reminder is due, and from proceeding to \textit{make booking} until payment is received. Five constrained duties and three constrained permissions are also correctly extracted within the XOR branches conditioned on recourse possibility.

The two \textit{Credit Scoring} models are the primary test cases for multi-party extraction and event-based gateway handling. Both models produce correct party assignments, correct duty and permission classifications, and correct wait prohibitions on the bank's downstream tasks. The asynchronous variant additionally exercises the event-based gateway, producing three wait prohibitions with distinct lifting conditions (\textit{credit score received} and \textit{delay information received}) and five permissions qualified by gateway conditions. Cross-role prohibitions are correctly emitted for both parties in both models; Figure~\ref{fig:extraction} illustrates the complete extracted policy for the synchronous case.

\textit{Self-Service Restaurant} is the most structurally complex model in the corpus, with three participants, ten message flows, five intermediate catch events, and one cycle. The pipeline detects and collapses the cycle into a meta-node, annotated as a loop over \textit{Call guest}, and proceeds without error. All 18 tasks across the three pools are correctly classified as unconditional duties, reflecting a process with no optional paths. The five wait prohibitions are all correctly attributed: the employee is prohibited from proceeding until the meal is ready, the guest is prohibited from placing an order until their turn arrives and from collecting their meal until notified, and the employee is prohibited from handing over the meal until the guest appears. Three compact cross-role prohibition rules correctly partition the task space across the three participants. The collapsed cycle is the one construct not faithfully represented in the output; it is emitted as an unconditional duty on the meta-node, losing the iterative semantics, which we discuss in Section~\ref{sec:limitations}.

Across all five models, every task is assigned to the correct party, every gateway condition is correctly qualified, and every intermediate catch event produces a wait prohibition with the correct lifting condition. The one structural limitation encountered is the cycle in the restaurant model, which is handled safely but not semantically faithfully, consistent with the known open problem discussed in the following section.

\section{Limitations and Open Problems}
\label{sec:limitations}

The pipeline handles the core BPMN vocabulary correctly across the five benchmark models, but four limitations are worth stating precisely.

The first concerns cycles. When Tarjan's algorithm collapses a strongly connected component into a meta-node, the iterative structure of the original subgraph is lost. The collapsed node is emitted as an unconditional duty, which is safe but semantically impoverished: it does not distinguish a task performed once from a task performed repeatedly while a condition holds. 
Resolving it faithfully would require either an ODRL profile extension for iteration semantics or a preprocessing step that unfolds statically bounded loops into conditional branches before traversal.

The second concerns unnamed pools. BPMN does not require pools to carry a name attribute, and several models in the wild, including the credit scoring benchmark, omit pool names entirely. Where a pool name is absent, the pipeline falls back to the raw XML participant identifier as the \texttt{odrl:Party} label, producing strings such as \texttt{Participant\_1x9bm3q} in the policy output. The structural correctness of the extracted rules is unaffected, but the policy becomes unreadable without reference to the source model. A fallback strategy using lane names or process names as substitutes is straightforward and is deferred to the next version.

The third concerns AND and OR gateways. The current implementation restricts the branch-local classification pass to XOR and event-based gateways; AND and OR branches are handled by the global reachability check alone, which is correct in the majority of cases but does not guarantee accurate ConstrainedDuty detection within complex parallel structures. The inclusive OR gateway carries an additional difficulty: its state-dependent merge semantics cannot be determined from local structure alone, and its deontic interpretation remains an open research question regardless of implementation.

The fourth concerns expressiveness boundaries in core ODRL. Temporal sequencing, the requirement that obligation A complete before obligation B may begin, is partially captured through the dominance tree but has no native ODRL expression. Similarly, two duties generated from an AND split are emitted as independent \texttt{odrl:Duty} rules, losing the requirement that they execute concurrently. Both cases are handled correctly within what core ODRL can express; the gap is in the target language, not the extraction, and points toward a profile extension as the appropriate resolution.

\section{Conclusion}
\label{sec:conclusion}

Policy infrastructure needs ODRL. Organisations have BPMN. The gap between them is not a gap in content but a gap in representation: the obligations, permissions, and prohibitions that ODRL policies express are already encoded in process models, structurally, by virtue of how BPMN control flow works. What has been missing is a principled, automatic route from one representation to the other.

The pipeline presented here provides that route. It is grounded in a formal structural correspondence, implemented as a graph traversal over BPMN XML, and produces valid ODRL policies in JSON-LD without requiring any policy authoring expertise from the process modeller. Evaluated on five benchmark models covering single and multi-party processes, optional and mandatory paths, event-based synchronisation, and cyclic structure, it correctly extracts duties, permissions, and prohibitions in all cases within the expressive reach of core ODRL.

The novel contribution beyond prior work is the treatment of intermediate catch events as \texttt{odrl:Prohibition} rules with lifting constraints, capturing waiting semantics that earlier approaches discarded. This is not a minor implementation detail. It reflects a principled reading of what these events mean normatively, and it makes the extracted policies semantically faithful in a way that bolted-on constraints cannot achieve.

The open problems identified in Section~\ref{sec:limitations} define the agenda for the future version of this work: a formal proof of the mapping's soundness, an ODRL profile addressing loops and temporal sequencing, systematic performance characterisation as a function of process size, and evaluation at scale across the participant-submitted model variants in the Camunda bpmn-for-research repository, which contains over 3,700 diagrams across the same four scenarios~\cite{camunda2015bpmnresearch, compagnucci2021trends}. The present paper establishes the foundation on which that work rests.

\section*{Acknowledgements}
 This work has been supported by HARNESS, funded from the EU's Horizon Europe research and innovation programme under grant agreement no. 101169409, and by MALTA PID2024-159504OB-I00 funded by MICIU/AEI/10.13039/501100011033.

 \section*{Declaration on Generative AI}
 During the preparation of this work, the authors used Claude Sonnet (Anthropic) to generate boilerplate code and assist with infrastructure setup, refine English, and correct grammar; all research decisions, experimental design, and results are solely the authors' own work.

\bibliography{references}

\end{document}